\documentclass[aps,prb,preprint,superscriptaddress,showpacs]{revtex4}
\usepackage{graphics}
\usepackage{graphicx}
\usepackage{amsmath}
\usepackage{amssymb}
\usepackage{bm}
\usepackage{dcolumn}

\begin{document}

   \title[]{Optical circular dichroism of single-wall carbon nanotubes}

 \author{Ariadna S\'anchez-Castillo}
 \affiliation{Instituto de F\'{\i}sica, Universidad Aut\'onoma de Puebla, Apartado Postal J-48, Puebla 72570,  M\'exico}

\author{C. E. Rom\'an-Vel\'azquez}
\affiliation{Instituto de F\'{\i}sica, Universidad Nacional Aut\'onoma
de M\'exico, Apartado Postal 20-364, D.F. 01000,  M\'exico}

\author{Cecilia Noguez}
\email[Corresponding author. Email:]{cecilia@fisica.unam.mx}
\affiliation{Instituto de F\'{\i}sica, Universidad Nacional Aut\'onoma
 de M\'exico, Apartado Postal 20-364, D.F. 01000,  M\'exico}

 \author{L. Meza-Montes}
 \affiliation{Instituto de F\'{\i}sica, Universidad Aut\'onoma de Puebla, Apartado Postal J-48, Puebla 72570,  M\'exico}

   \date{\today}

 \begin{abstract}
The circular dichroism (CD) spectra of single-wall carbon nanotubes are calculated using a dipole approximation. The calculated CD spectra show features that allow us to distinguish between nanotubes with different angles of chirality, and diameters. These results provide theoretical support for the quantification of chirality and its measurement, using the CD lineshapes of chiral nanotubes. It is expected that this information would be useful to motivate further experimental studies.
 \end{abstract}


 \maketitle

\section{Introduction}

A century ago Lord Kelvin stated, "I call any geometrical figure, or group of points, chiral, and say that it has chirality, if its image in a plane mirror, ideally realized, cannot be brought to coincide with itself." According to Kelvin's definition, one could say only whether an object is chiral or not (achiral), and infering that chirality is a purely geometrical property there is not reason to relate it to chemistry, physics or biology. Decades before, Louis Pasteur discovered a connection between optical activity and molecular chirality, and found that substances with the same elementary composition have different physical properties that led him to suppose that forces of nature are not mirror-symmetric. Today, we know that chirality plays an important role in chemistry, physics and biology and that left-amino acids and left-peptides are predominant in the living world.~\cite{barron,rikken}
Despite of the simplicity of Kelvin's definition of chirality, there is not an algorithm for such a criterion to  diagnostic chirality. \cite{smirnov,bellarosa,buda}  
There are  several works which have attempted to estimate chirality quantitatively,  however, they have not succeeded to find an universal approach which gives unambiguous results.
\cite{richter,damhus,kut,zabro,moreau,maruani,thomas,nogradi,sokolov}
Furthermore, a drawback of these attempts to quantify chirality is that they do not provide a way to compare directly with experimental observations.

Among nanostructures, carbon nanotubes are known to be chiral. The atomic structure of single-wall carbon nanotubes (SWNTs) resembles the wrapping of a sheet of carbons located in a two-dimensional hexagonal lattice to form a cylinder, see Fig.~\ref{f1}. The sheet can be rolled up in different ways, such that, nanotubes with similar diameters have different chirality. Despite of the fact that the atomic structure of SWNTs is simple, their properties depend dramatically on chirality.~\cite{saito} SWNTs can be described with a chiral vector $\vec{C}_h$ given by the unit vectors of the hexagonal lattice $\vec{a}_1$ and $\vec{a}_2$, as
$
\vec{C}_h = n \vec{a}_1 + m \vec{a}_2,
$
where $n$ and $m$ are integers. SWNTs are frequently  denoted using these integer number as the nanotube $(n,m)$. The diameter of the nanotube is $d = s/\pi$ where $s$ is the circumferential length of the nanotube, $s = |\vec{C}_h | = a \sqrt{n^2 + m^2 + nm}$, with $a$ the lattice constant. We can define also the chiral angle $\theta_c$ as the tilt angle of the hexagons with respect to the direction of the SWNTs axis:
\begin{equation}
\label{angle}
\cos{\theta_c} = \frac{\vec{C}_h \cdot \vec{a}_1}{|\vec{C}_h | | \vec{a}_1 |} = \frac{2n + m}{2\sqrt{n^2 + m^2 + nm}}\, .
\end{equation}
There are only two classes of achiral SWNTs: the armchair with $n=m$ and $\theta_c=30^\circ$, and the zigzag nanotube with $n\neq m$, $m=0$ and $\theta_c = 0^\circ$. SWNTs with $ 0^\circ < \theta_c < 30^\circ$ are all chiral.  


Recently, it has been shown that linear and circular dichroisms can be  useful techniques to study SWNTs.~\cite{rajendra,tasaki,Ge} Quantum-mechanical calculations of the influence of chirality on the optical properties of small SWNTs~\cite{tasaki,Ge,chang,marino} had been performed. Tasaki \textit{et al.},~\cite{tasaki} calculated the optical properties of SWNTs using a tight-binding approach. They found that chiral nanotubes are optically active,  the CD spectra oscillates, and CD decreases as the diameter increases. However, they did not find a relation between the chiral angle and the strength of CD. Recently, Samsoninidze and collaborators~\cite{Ge} also employed a tight binding approach  to calculate the electron transitions for chiral SWNTs and circularly polarized light propagating along the nanotube axis. They found the selection rules  of the electron transition depending on the handeness of SWNTs that gives rise to optical activity when the time-reversal symmetry is broken, yielding to CD. They calculated the optical absorption for the (20,10) SWNT for left- and right-circularly polarized light. From these spectra, we can infer that they did not find a oscillatory behavior of CD, as Tasaki \textit{et al.},~\cite{tasaki} did. In both calculations, the tight-binding approximation only accounts for $\pi$-band electrons, and other important interactions and effects, like many-body and local field, are not included. In particular, it was found from \textit{ab initio} calculations that many-body~\cite{chang} and local field~\cite{marino} effects play important roles to determine the optical properties of nanotubes. Furthermore, local field effect has been pointed out as the main source of the induced optical anisotropy at surfaces of cubic crystals.~\cite{barrera,mochan,hogan} However, quantum mechanical approaches are still inappropriate for systems with hundreds and thousands of atoms due to the huge computational effort that is involved in these type of calculations.  

Our goal in this work is to obtain a way to quantify chirality for SWNTs by finding a relation between the chiral angle of SWNTs and their circular dichroism spectra.  To explore the capability of the circular dichroism (CD) tool to detect different angles of chirality existing in nanotubes it would be useful to have a theoretical estimation of their CD behavior.  We employ a classical electromagnetic model to simulate CD spectra of chiral carbon nanotubes, which includes in an approximated way many body and local field effects. We consider the nanotubes  as systems of coupled point dipoles, where the atomic polarizability are obtained from the dielectric function of graphite. The classical approach employs an ansatz in which a localized polarizable unit in the system in the presence of an external field responds to the local field due to all other induced dipole moments at the other sites plus the external field. This method can treat systems of thousands of atoms for a relatively low cost.  It is expected that this information would be useful to motivate further experimental studies to correlate distinctive features of the CD spectra with the different angles of chirality and diameters of SWNTs.

\section{Dipole Approximation}

To study the optical response of medium size nanotubes we employ the dipole approximation. This approximation has been recently used to study the optical response of chiral gold nanoclusters~\cite{roman}, as well as the electron energy loss spectra of SWNTs~\cite{rivacoba,wu} and fullerenes.~\cite{henrard}

Let us assume that the SWNTs of interest are composed by $N$ carbon atoms represented by a polarizable point dipole located at the
position of the atom. We assume that the dipole located at
$\mathbf{r}_{i}$, with $ i=1,2, \dots ,N$, is characterized by a
polarizability $\boldsymbol{\alpha}_{i} (\omega)$, where $\omega$
denotes the angular frequency. The SWNT is excited by an incident
circular polarized wave with wave-vector parallel to the axis of the nanotube.
Each dipole of the system is subjected to a total electric field
which can be divided into two contributions: (i) $\mathbf{E}_{i,\mathrm{inc}}$, the incident radiation field,  plus (ii) $\mathbf{E}_{i,\mathrm{dip}}$, the radiation field resulting from
all of the other induced dipoles. The sum of both fields is
the so called local field given by
\begin{equation}
\mathbf{E}_{i,\mathrm{loc}} = \mathbf{E}_{i,\mathrm{inc}} +
\mathbf{E}_{i,\mathrm{dip}} = \mathbf{E}_{i,\mathrm{inc}} -
\sum_{i\neq j} \mathbb{T}_{ij} \cdot \mathbf{p}_{j}\,,
\label{local}
\end{equation}
where $\mathbf{p}_{i}$ is the dipole moment of the atom
located at $\mathbf{r}_{i}$, and $\mathbb{T}_{ij}$ is an
off-diagonal matrix which couples the interaction between dipoles.~\cite{purcell,draine} 
On the other hand, the induced dipole moment at each atom is given by
$\mathbf{p}_{i}=\boldsymbol{\alpha}_{i}\cdot \mathbf{E}_{i,\mathrm{loc}}$, such that $3N$-coupled complex linear equations are obtained from Eq.~(\ref{local}). These equations can  be rewritten as
\begin{equation}
\sum_{j}\left[(\boldsymbol{\alpha}_{j})^{-1} \delta_{ij} +
 \mathbb{T}_{ij} (1 - \delta_{ij}) \right] \cdot \mathbf{p}_{j} = 
\mathbb{M}_{ij} \cdot \mathbf{p}_{j} =  \mathbf{E}_{i,\mathrm{inc}} , 
 \label{pes}\end{equation}
where the matrix $\mathbb{M}$ is composed by a diagonal part given
by $(\boldsymbol{\alpha}_{j})^{-1} \delta_{ij}$ and by an
off-diagonal part given by the interaction matrix 
\begin{eqnarray}
\mathbb{T}_{ij} \cdot \mathbf{p}_{j} &=& \frac{e^{ikr_{ij}}}
{r_{ij}^{3}} \Bigg\{k^{2}
\mathbf{r}_{ij} \times (\mathbf{r}_{ij}\times \mathbf{p}_{j}) \\
&+& \frac{(1-ikr_{ij})} {r_{ij}^{2}}\left[ r_{ij}^{2}
\mathbf{p}_{j}-3 \mathbf{r}_{ij} (\mathbf{r}_{ij}\cdot
\mathbf{p}_{j})\right] \Bigg\}.  \notag
\end{eqnarray}
Here $\mathbf{r}_{ij} =\mathbf{r}_{i} - \mathbf{r}_{j}$,  $r_{ij}=|\mathbf{r}_{ij}|,$
and $k$ is the magnitude of the wave vector of the incident electromagnetic field.

The diagonal part in Eqs.~(\ref{pes}) is related to the polarizability
of each atom,  i. e. to the material properties of the system, while the off-diagonal part depends only  on the atomic positions, i.e. to 
the geometrical properties. 
Once we solve the complex-linear equations shown in Eq. (\ref{pes}),
the dipole moment on each atom in the nanotube can be determined, and then we can calculate the
extinction  cross section, $C_{\mathrm{ext}}$ of the SWNT.
In terms of the dipole moments, \cite{purcell,draine}
\begin{equation}
C_{\mathrm{ext}} =\frac{4\pi k}{{E}_{0}^{2}} \sum_{i=1}^{N}
\mathrm{Im} (\mathbf{E}_{i,\mathrm{inc}}^{\ast }\cdot \mathbf{p}_{i}) \, ,
\end{equation}
where ($\ast $) means complex conjugate. 

Circular dichroism is defined as the differential absorption of left (L) and right (R) hand circularly polarized light, which is obtained by subtracting the corresponding extinction efficiencies, $Q_{\mathrm{ext}}$,  as
\begin{equation}
{\mathrm CD} = Q_{\mathrm{ext}}^{\text{R}} - Q_{\mathrm{ext}}^{\text{L}} . \label{cd}
\end{equation}
Here, the extinction efficiency is defined as $Q_{\mathrm{ext}} = C_{\mathrm{ext}}/A$, where $A =  \pi s L$, and $L$ is the length of the unit cell of the nanotube. Edge effects due to the finite length of the nanotube are removed by using periodic boundary conditions. 

Because chirality is a geometrical property,
the CD spectrum will be more sensitive to the second term inside
the bracket in Eq.~(\ref{pes}), which depends on the atomic geometry,
than to the first one that is related to the atomic polarizability.
Therefore, the simulated CD spectra will include the geometrical
information about the chiral SWNTs. Moreover, it will be crucial for the calculations to have ``realistic'' atomic positions to obtain the right curvature for each one of the SWNTs studied here.

In this work, we assume that the polarizability for each atom $\boldsymbol{\alpha}_i$ is isotropic and 
is the same for all the atoms in the nanotube ($\boldsymbol{\alpha}_{i} = \alpha_i \equiv  \alpha$). We obtained the polarizability from  the Clausius-Mossotti relation, where the dielectric function of the system was taken from the experimental data for graphite.~\cite{grafito}
This approximation has been used before to study carbon nanotubes
providing  good qualitative agreement between calculated optical
properties and experimental data.~\cite{henrard,wu,rivacoba}

\section{Atomic structures}

We obtained the atomic structures of the SWNTs  within the density functional theory (DFT) using the \textsc{siesta} computer code.~\cite{siesta} This code has been widely employed to study the atomic relaxation and electronic properties of small SWNTs.~\cite{ordejon1,ordejon2,ordejon3,ordejon4} We 
used Perdew-Burke-Erznerhof exchange correlation functional \cite{Perdew}
within the generalized gradient approximation (GGA). To take into account the interaction between valence electrons and ionic cores, we employed fully non-local norm-conserving  pseudopotentials proposed by Troullier and Martins.~\cite{Martins} A double $\zeta$ polarized (DZP) basis set was used with cutoff radii of 5.12 and 6.25 atomic units for the $2s$ and $2p$ orbitals, respectively. In this work, we consider that SWNTs are infinitely long by repeting the unit cell of length $L$ along the nanotube axis. More details of the calculation can be find elsewhere.~\cite{ordejon1,ordejon2,ordejon3,ordejon4}

\begin{figure} [h]
\begin{center}
\includegraphics{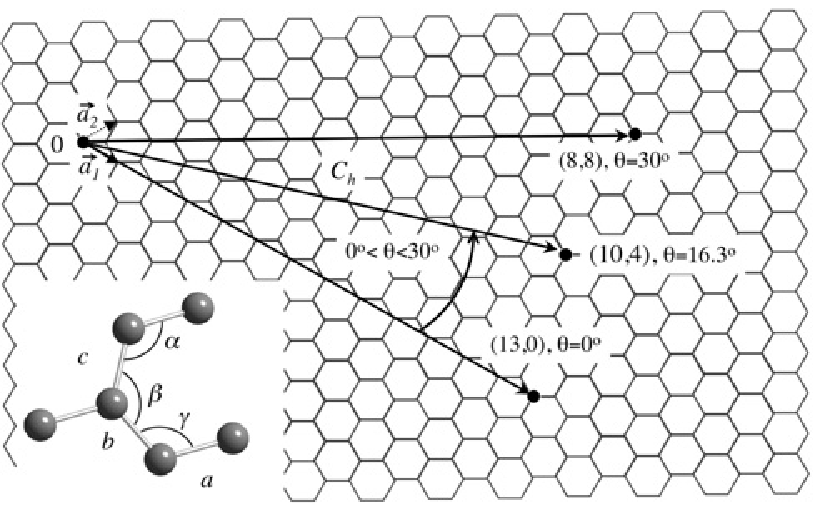} 
\end{center} 
\vskip-.3cm
\caption{Model of the atomic geometry of a hexagonal two-dimensional lattice with unit vectors $\vec{a}_1$ and $\vec{a}_2$.}
\label{f1}
\end{figure}

In Table I, we summarized the geometrical parameters obtained  in this work using \textsc{siesta}, as labeled in Fig.~\ref{f1}. For comparison, we also include the corresponding parameters of the non-relaxed SWNTs. After relaxation, we found that the chiral vector and diameter of the nanotubes increase by less than 2\%
with respect to their corresponding non-relaxed (n. r.) values. The  bond length in the radial direction,
 denoted by $a$ in Fig.~\ref{f1}, changes more than the bond lengths in the directions parallel to the axis of SWNTs. These bonds are denoted by $b$ and $c$ in Fig.~\ref{f1}.  For SWNTs with diameters about 
1~nm, the bond lengths $a$ and $c$ are similar, and they are always larger than $b$. For SWNTs with larger diameters, the bond lengths $a$, $b$, and $c$ become almost equal. The unit cell length, $L$,  is also deviated from its non-relaxed value, increasing approximately 1\% in all cases. Then, the bond angles (see Fig.~\ref{f1}) are also modified upon relaxation, such that the $\alpha$ angle is larger, while the $\beta$ and $\gamma$ angles tend to be reduced. Our results of the atomic geometry of SWNTs are in excellent agreement with previous DFT calculations for SWNTs with small diameters~\cite{ordejon1,ordejon2,ordejon3,ordejon4}.

\section{Circular dichroism spectra} 

We present results for the extinction efficiency and circular dichroism spectrum of different SWNTs. The spectra were calculated using the dipole approximation introduced in  Section~II, and the atomic positions of SWNTs obtained in Section III.
We first present results for the extinction efficiency of SWNTs as a function of the frequency. Then, we discuss the CD spectra for different SWNTs with the same diameter but different chirality, and finally the CD spectra for SWNTs with the same chirality but different diameter. 

\begin{figure} [tbh]
\centerline {
\includegraphics{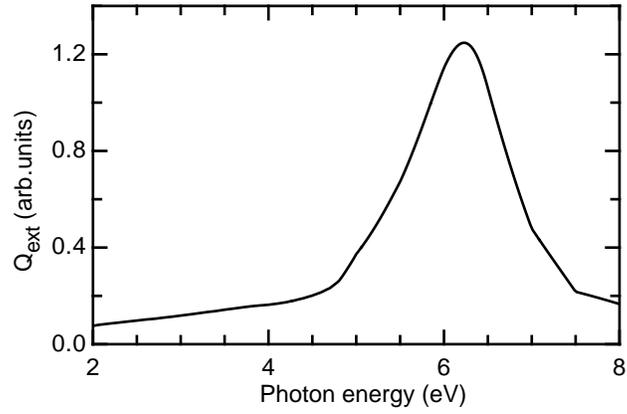} 
}
\caption{Extinction efficiency of the SWNT  (13,1).}
\label{f2}
\end{figure}

In Fig. ~\ref{f2} we show the extinction efficiency $Q_{\text{ext}}$ in arbitrary units (a. u.) of the SWNT (13,1), as a function of the photon energy of the incident light  from 2~eV (620~nm) to 8~eV (155~nm).  We observe that $Q_{\rm ext}$ has a maximum at 6.2~eV where absorption effects are dominant, while the asymmetry of the peak is due to scattering effects because the SWNTs considered here are infinitely long. Therefore,  a chiral SWNT will show  dichroism around these energies. We only show the $Q_{\rm ext}$ for one SWNT, however the same behavior is observed for the rest of the nanotubes.

\begin{figure}[tbh] 
\centerline {
   \includegraphics{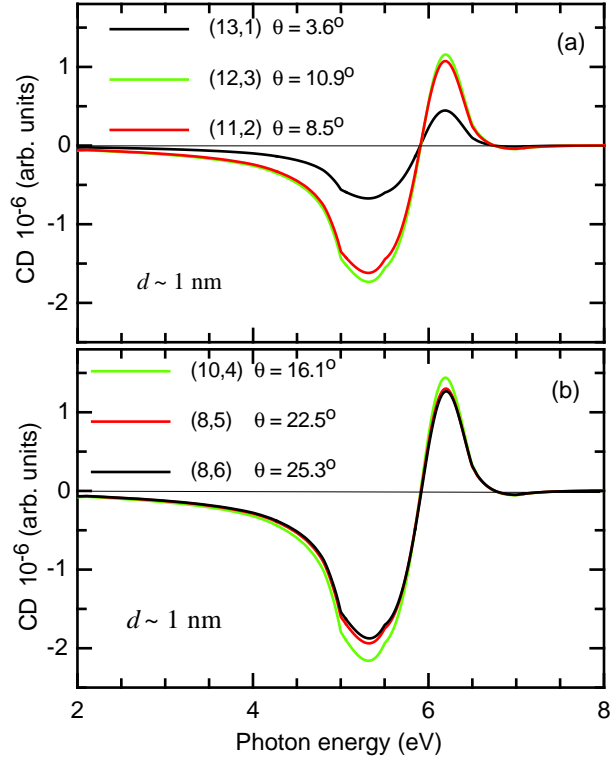} 
   }
   \caption{(Color online) CD spectra of SWNTs with the same diameter $d =1$~nm and (a) $0^\circ \leqslant \theta_c \leqslant 15^\circ$, and (b) $15^\circ \leqslant \theta_c \leqslant 30^\circ$.}
   \label{f3}
\end{figure}
\begin{figure} [tbh]
\centerline {
\includegraphics{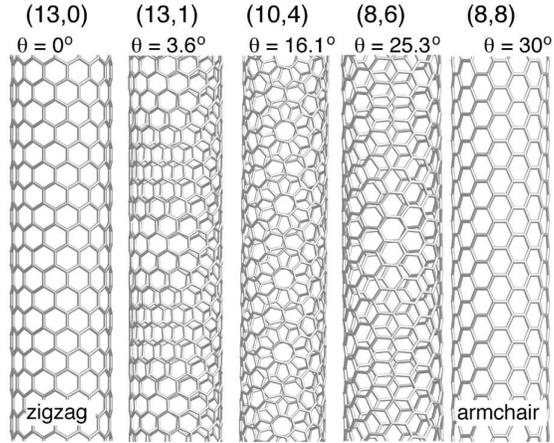} 
}
\caption{Structure of single-wall SWNTs  with different chiral angles between $0^\circ$ and $30^\circ$.}
\label{f4}
\end{figure}

In Fig.~\ref{f3}  we present the CD spectra for single-wall SWNTs of about the same diameter,  $d =1$~nm, but different chiral angle $\theta_c$.  We show the CD in arbitrary units (a. u.) as a function of the photon energy  of the incident light from 2~eV  to 8~eV. All the spectra show a minimum at 5.2~eV (238~nm) and a maximum at 6.1~eV (203~nm), where light absorption of graphite is more intense.~\cite{grafito} From Fig.~\ref{f3}(a), we observe that the CD spectrum is more intense as the chiral angle in creases, while in Fig.~\ref{f3}(b) the contrary is observed. For achiral  nanotubes,  armchair ($\theta_c=30^\circ$) and zigzag ($\theta_c = 0^\circ$), the CD spectra is always null.   We found that the maximum of the CD spectra is reached when the helicity of the nanotubes is also a maximum, i. e., when $\theta_c$ is close to  $15^\circ$, as shown in the nanotube structures in Fig.~\ref{f4}. For all diameters of SWNTs, we found  the same behavior of the CD spectrum as a function of t
he chiral angle.  In 
summary, we found that  the intensity of the CD spectrum depends on the chiral angle of the SWNTs. Therefore, we can conclude that CD measurements can be useful to quantify chirality in these nanostructures.

\begin{figure} [tbh]
\vskip.1cm
\centerline {
\includegraphics{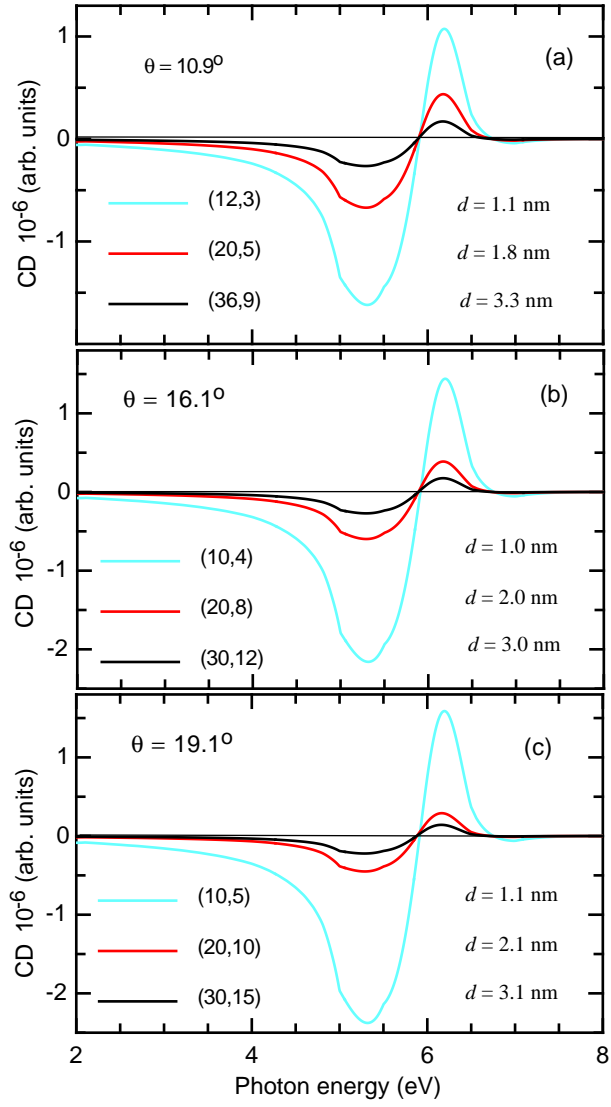} 
}
\caption{(Color online) CD spectra of  SWNTs of different diameter but same chiral angle (a) $\theta_c = 10.9^\circ$, (b) $\theta_c = 16.1^\circ$, and (c) $\theta_c = 19.1^\circ$.}
\label{f5}
\end{figure}

Now, we analize the CD spectra for nanotubes of different diameter but same chiral angle. In Fig.~\ref{f5} we show the CD spectra for nanotubes with (a) $\theta_c = 10.9^\circ$ and diameters $d = 1.1, \,\, 1.8,\,\,{\rm and } \,\, 3.3$~nm; with (b) $\theta_c = 16.1^\circ$ and diameters $d = 1.0, \,\, 2.0,\,\,{\rm and } \,\, 3.0$~nm, and with (c) $\theta_c = 19.1^\circ$ and diameters $d = 1.1, \,\, 2.1,\,\,{\rm and } \,\, 3.1$~nm. For the three cases, we found that the intensity of the CD spectrum is always larger for smaller diameters. We observe that the maximum of the CD spectra is about half (one third) when the diameters is twice (three times) larger. This is due to the fact that wider nanotubes with the same helicity have a larger surface, such that the chiral effect is less intense. In the limit case, when the diameter goes to infinity, and we  recover the hexagonal sheet, the CD will be zero even when $\theta_c$ will remain unchanged. In summary, we found that there is a relation between the intensity of the CD spectra with the diameter of the nanotube. From these results, we can conclude that CD measurements can be also useful to quantify the diamater of chiral SWNTs.


\section{Conclusions}

Using the  computer code called \textsc{siesta}, within the density functional theory, we obtained the atomic positions of 14 different single-wall carbon nanotubes such that most of them were not reported before in the literature. Our DFT results are in excellent agreement with previous calculations of smaller single-wall carbon nanotubes.  Furthermore, we have studied the circular dichroism optical spectra of these single-wall carbon nanotubes of different diameters and chirality within a dipole approximation. We found that for nanotubes with a given diameter,  the intensity of circular dichroism spectra increases as the helicity of the nanotube also does. We also found that the intensity of the circular dichroism spectra of single-wall nanotubes, for a given chiral angle, depends on the diameter of the nanotube, such that wider nanotubes show a less intense spectra. From these results,  we can conclude that circular dichroism measurements can be useful to quantify chirality as well the diameters of chiral SWNTs. It is expected that this information would be useful to motivate further experimental studies in this direction.

 \acknowledgments
 We acknowledge the fruitful discussion with Ignacio L. Garz\'on. Partial financial support from DGAPA-UNAM grant No.~IN101605, and VIEP-BUAP II 193-04/EXC/G is also acknowledged. 


\newpage

\newpage

\begin{table} [h]
\begin{center}
\begin{tabular}{|c|c|c|c|c|c|c|c|c|c|}
\hline \hline
$(n,m)$&$\theta_c$&$|C_h|$(\AA)&$a$(\AA)&$b$(\AA)&$c$(\AA)&$\alpha$&$\beta$&$\gamma$ \\
\hline
(13,1)& 3.6       & 34.46  &  1.46 & 1.45 & 1.46 & 118.8 & 120.0 & 119.9 \\
n. r. &   & 33.74  & 1.44 & 1.44 & 1.44 & 118.6 & 120.1 & 120.0 \\ \hline 
(26,2)&	3.6 & 10.87 & 1.45 & 1.45 & 1.46 & 119.7 & 120.0 & 120.0 \\	
n. r. &  & 10.74 & 1.44 & 1.44  & 1.44 & 119.6 & 120.0 & 120.0 \\ \hline
(11,2) & 8.5      & 30.76   & 1.46 & 1.45 & 1.46 & 118.5 & 120.0 & 119.8 \\
n. r. &   & 30.24  &  1.44 & 1.44 & 1.44 & 118.3 & 119.9 & 120.1 \\ \hline
(12,3) & 10.9    & 34.80   &  1.46 & 1.46 & 1.46 & 118.9 & 119.9 & 119.8 \\
n. r. &   & 34.29   &  1.44 & 1.44 & 1.44 & 118.7 & 120.1 & 119.9 \\ \hline
(20,5) & 10.9    & 57.84   & 1.46 & 1.45 & 1.46 & 119.6 & 120.0 & 119.9 \\
n. r. &   & 57.11   & 1.44 & 1.44 & 1.44 & 119.5 & 120.0 & 119.5 \\ \hline
(36,9) & 10.9    & 103.95 &  1.45 & 1.45 & 1.46 & 119.9 & 120.0 & 120.0 \\
n. r. &  & 102.87 &  1.44 & 1.44 & 1.44 & 119.8 & 120.0 & 120.0 \\ \hline
(10,4) & 16.1    & 31.67  &  1.46 & 1.45 & 1.46 & 118.8 & 119.8 & 119.7 \\
n. r. &   & 31.15  &  1.44 & 1.43 & 1.44 & 118.6 & 120.1 & 119.7 \\ \hline
(20,8)	& 16.1	& 63.05	& 1.46	& 1.45	& 1.46	& 119.7	& 120.0	& 119.9	\\
n. r.	& & 62.30 & 1.44 & 1.44	& 1.44	& 119.7	& 120.0	& 119.9	\\ \hline
(30,12) & 16.1 & 63.99 & 1.47 & 1.47 & 1.47 & 119.9 & 120.0 & 120.0 \\
n. r.	& & 56.50 & 1.44 & 1.44	& 1.44	& 119.8	& 120.0	& 120.0	\\ \hline
(10,5) & 19.1    & 33.52  &  1.46 & 1.45 & 1.46 & 119.0 & 119.8 & 119.7 \\
n. r. &   & 32.99  & 1.44 & 1.43 & 1.44 & 118.9 & 120.1 & 119.7 \\ \hline
(20,10) & 19.1	& 66.76	& 1.46	& 1.46	& 1.46	& 119.8	& 120.0	& 119.9	\\
n. r.	& & 65.97 & 1.44 & 1.44	& 1.44	& 119.7	& 120.0	& 119.9	\\ \hline
(30,15)	& 19.1	& 100.03 & 1.45	& 1.45	& 1.46	& 119.9	& 120.0	& 120.0	\\
n. r.	& & 98.96 & 1.44 & 1.44	& 1.44	& 119.9	& 120.0	& 120.0	\\ \hline
(8,5) & 22.4      & 28.84  &  1.46 & 1.45 & 1.46 & 118.9 & 119.6 & 119.5 \\
n. r. &   & 28.33  &  1.44 & 1.43 & 1.44 & 118.6 & 119.4 & 120.0 \\ \hline
(8,6) & 25.3      & 30.85  &  1.46 & 1.45 & 1.46 & 119.1 & 119.6 & 119.6 \\
n. r. &   & 30.34  & 1.44 & 1.43 & 1.44 & 118.9 & 119.4 & 120.0 \\ \hline

\hline \hline
\end{tabular}
\end{center} \vskip-.5cm
\caption
{Geometrical parameters as defined in Fig.~1 for relaxed and non-relaxed (n. r.)  SWNTs.  $\theta_c$, $\alpha$, $\beta$ and $\gamma$ are given in degrees.}
\label{parameters}
\end{table}

\end{document}